\documentclass[runningheads]{llncs}
\usepackage{amsmath}
\numberwithin{figure}{section}
\numberwithin{table}{section}
\numberwithin{equation}{section}
\usepackage{graphicx}
\usepackage{multirow}

\begin{document}
\title{A multi-view multi-stage and multi-window framework for pulmonary artery segmentation from CT scans}

\author{ZeYu Liu\inst{1} \and
Yi Wang\inst{2} \and  Jing Wen \inst{3} \and 
Yong Zhang\inst{4}
\and 
Hao Yin\inst{5}
\and
Chao Guo\inst{6} \and 
ZhongYu Wang\inst{7}
}
\authorrunning{ZeYu Liu et al.}

\institute{Chongqing University, Chongqing city, China \email{cvlearn@163.com} \and Chongqing University, Chongqing city, China \email{yiwang@cqu.edu.cn} \and Chongqing University, Chongqing city, China \email{wj@cqu.edu.cn}
\and
Chongqing University, Chongqing city, China \email{zhangyong7630@163.com} \and
Chongqing University, Chongqing city, China \email{202114131132@cqu.edu.cn} \and
Chongqing University, Chongqing city, China \email{mzlxavier1230@gmail.com} \and
Ziwei king star Digital Technology Co., Ltd, Hefei city China \email{wzhongyu@ziweidixing.com} 
}
\maketitle              
\begin{abstract}
This is the technical report of the 9th place in the final result of PARSE2022 Challenge. We solve the segmentation problem of the pulmonary artery by using a two-stage method based on a 3D CNN network. The coarse model is used to locate the ROI, and the fine  model is used to refine the segmentation result. In addition, in order to improve the segmentation performance, we adopt multi-view and multi-window level method, at the same time we employ a fine-tune strategy to mitigate the impact of inconsistent labeling.

\end{abstract}
\section{Introduction}
The segmentation of the pulmonary artery is of great significance, however performing a one-stage segmentation makes the results less accurate. For this we implemented a two-stage method for pulmonary artery segmentation. The advantage of the two-stage method is high accuracy. Manual annotation is often performed at a certain plane, In order to allow the network to obtain the comprehensive information of the three planes of the patch, we use a multi-view training and inference method. In order to highlight the tissue of the segmentation target, we adopt a special multi-window method to distinguish trunk and branches.
Compared with 2D CNN, 3D CNN can obtain more one-dimensional information, so we design a novel 3D CNN network, and the details  is in our upcoming published paper. To address the inconsistency of manual annotations, we employ a fine-tuning strategy to modify the inference results at the testing phase.

We summarized our work as below:
(1) A multi-stage, multi-view, multi-window framework for pulmonary artery segmentation.
(2) A novel 3D CNN network.
(3) A Fine-tune strategy to mitigate the impact of inconsistent labeling.

\section{Method}
\subsection{Preprocessing}
We adopted the following data preprocessing method.
As for coarse model, we used the method like below: 
    \begin{itemize}
        \item[$\bullet$] scale intensity
        \item[$\bullet$] resize
    \end{itemize}
As for fine model, we use the preprocessing method like below:
    \begin{itemize}
        \item[$\bullet$] scale intensity
        \item[$\bullet$] rand crop 
    \end{itemize}

We notice that the main pulmonary artery and the branches are in different
intensity distribution intervals. To distinguish between these two different types of tissues, we use two window levels and do not truncate the intensity in order to preserve the information. Therefore, both the coarse model and the fine model have two input channels, as shown in Fig 2.1, one channel has intensity value distribution [-900, 0], and the other is [0, 300].

\begin{figure}
\centering
\includegraphics[width=0.6\linewidth]{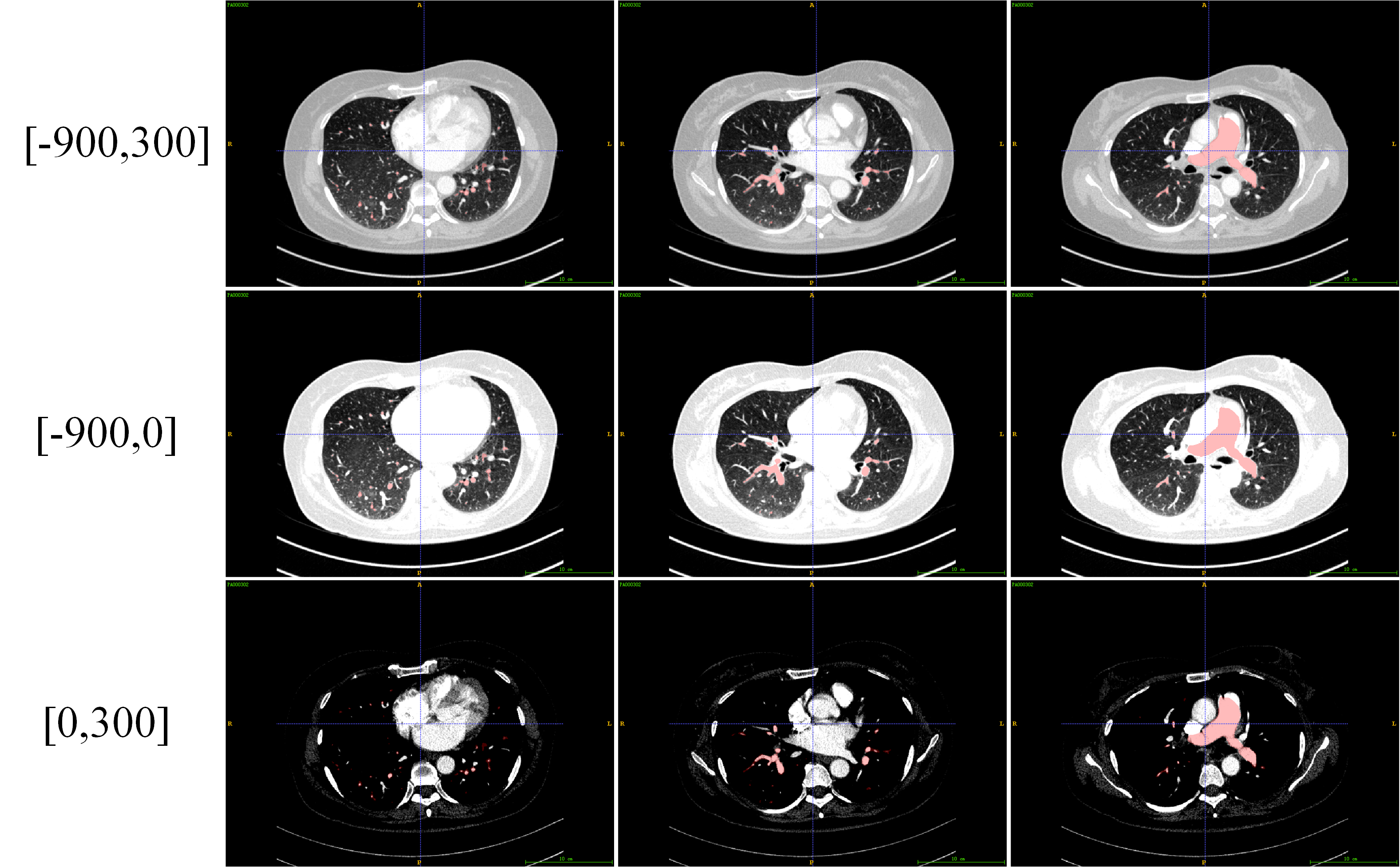}
\caption{Different window levels}\label{fig4}
\end{figure}

\subsection{proprosed method}
As for our proposed method, the first stage is used for locatig the ROI, and the second stage is used for refining segmentation result. The process is shown in at Fig 2.2. When training coarse model, the original image is scaled by trilinear interpolation, and the label is scaled by nearest neighbor interpolationo-fine is a common strategy in medical image segmentation, which has been studied for a long time.\cite{Selvarani2012MedicalIF,Wu2021CoarsetoFineLN,Zabihollahy2021FullyAM,Zhu2018A3C}

When training coarse model, the original image is scaled by trilinear interpolation, and the label is scaled by nearest neighbor interpolation. When training the coarse model, the original image is scaled by trilinear interpolation, and the label is scaled by nearest neighbor interpolation, the purpose of this is to accurately locate the pulmonary artery region. The training data of the second stage is randomly cropped according to the area determined by the label, and samples are separated into positive and negative based on the patch's center voxel. We randomly choose one of the coronal, sagittal, and horizontal planes as the orientation of the patch during each iteration of the fine model training.

\begin{figure}
\includegraphics[width=\textwidth]{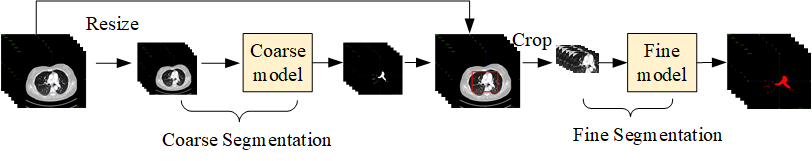}
\caption{The proposed solution flow, one stage is used to locate the pulmonary artery region, and the second stage is used for fine segmentation}\label{fig2}
\end{figure}

\subsection{Network}

In the field of image segmentation, U-Net\cite{Ronneberger2015UNetCN} is one of the most popular methods, as shown in Fig 2.3, for this challenge We also based on the idea of its classical structure to design our network. The left side of the network is an encoder and the right side is a decoder. Feature fusion is achieved by skip-connection between encoder and decoder. Because of the limited space and article type, the specific details of the network and other work will be in our upcoming published paper, so it is inconvenient to discuss details in this technical report. In this challenge, whether it is the first stage or the second stage, we both use our network to achieve our goal.

                                        \begin{figure}
                                        \centering
                                        \includegraphics[width=0.6\linewidth]{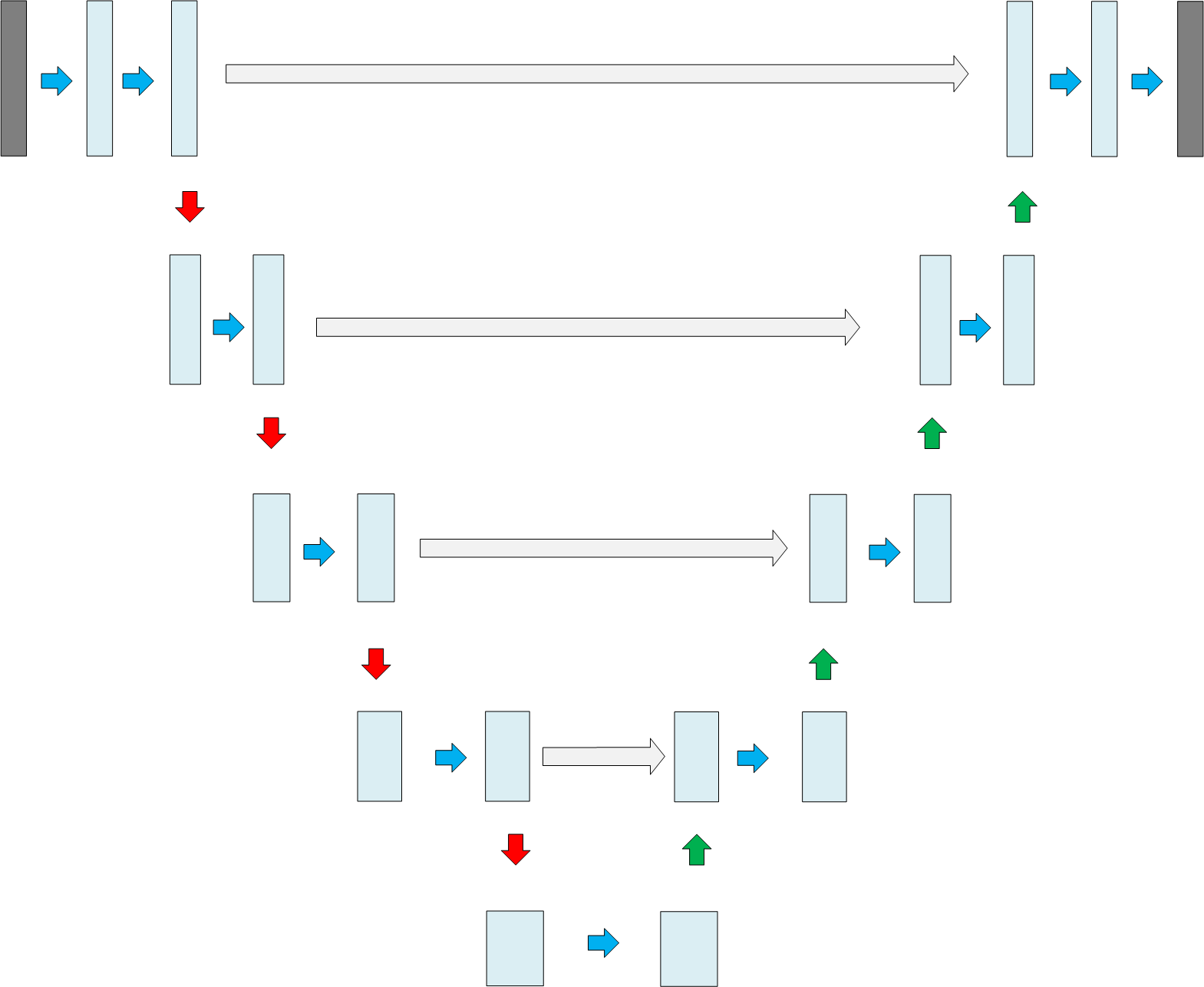}
                                        \caption{Classic 5-layer U-Net structure} \label{fig3.png}
                                        \end{figure}

\section{Dataset and Evaluation Metrics}
The challenge obtained 200 3D volumes from 10 doctors with refined pulmonary artery labeling, 100 for the training dataset, 70 for the closed testing dataset, and 30 for the opened validated dataset. As evaluation metrics, Multi-level Dice, Multi-level HD, and time-cost are used. In this competition, whether it is the first stage or the second stage, we both use this network to achieve our goal.

The Dice coefficient is a common set similarity measure function in semantic segmentation. It is usually used to calculate the similarity between two samples, and its value ranges from [0, 1].
                        \begin{equation}
                    Dice=\frac{2|X \bigcap Y|}{|X|+|Y|}
                        \end{equation}

In semantic segmentation, Dice is more sensitive to the inner filling of the label, while hausdorff distance is more sensitive to the boundary. The formula is defined as below.

                        \begin{equation}
                       H(A, B)=\max (h(A, B), h(B, A))
                        \end{equation}

                        \begin{equation}
                        h(A, B)=\max _{a \in A}{\min _{b \in B}\|a-b\|}
                        \end{equation}
                        
                        \begin{equation}
                        h(B, A)=\max _{b \in B}{\min _{a \in A}\|b-a\|}
                        \end{equation}

\section{Implementation Details}
Here are the details of our implemented environment and training process. In segmentation tasks, class imbalance is a very common phenomenon, like V-Net\cite{Milletari2016VNetFC} we also use Dice loss to solve this problem. Coarse model training parameter is set as Table 4.1, and Fine model training parameter is set as Table 4.2.


\begin{table}
\centering
\caption{coarse model training parameter}
\begin{tabular}{|l|l|} 
\hline
Initial method                                                                                   & nn.init.kaiming\_normal                                                                            \\ 
\hline
Batch size                                                                                       & 2                                                                                                  \\ 
\hline
Patch size                                                                                       & 192 192 192                                                                                        \\ 
\hline
Total epoches                                                                                    & 200                                                                                                \\ 
\hline
Optimizer                                                                                        & Novograd                                                                                           \\ 
\hline
Loss                                                                                             & DiceLoss                                                                                           \\ 
\hline
Initial learning rate                                                                            & 8.00E-03                                                                                           \\ 
\hline
\begin{tabular}[c]{@{}l@{}}Learning rate \\decay schedule\end{tabular}                           & LinearWarmupCosineAnnealingLR                                                                      \\ 
\hline
\begin{tabular}[c]{@{}l@{}}Stopping criteria, \\and optimal model selection criteria\end{tabular} & \begin{tabular}[c]{@{}l@{}}training 200 epoches, \\no other model selection criteria\end{tabular}  \\
\hline
\end{tabular}
\end{table}

\begin{table}
\centering
\caption{Fine model training parameter}
\begin{tabular}{|l|l|} 
\hline
Initial method                                                                                   & nn.init.kaiming\_normal                                                                                      \\ 
\hline
Batch size                                                                                       & 2                                                                                                            \\ 
\hline
Patch size                                                                                       & 192 192 192                                                                                                  \\ 
\hline
Each voxel case sample patch                                                                     & 4                                                                                                            \\ 
\hline
Total epoches                                                                                    & 300                                                                                                          \\ 
\hline
Optimizer                                                                                        & Novograd                                                                                                     \\ 
\hline
Loss                                                                                             & DiceLoss                                                                                                     \\ 
\hline
Initial learning rate                                                                            & 8.00E-03                                                                                                     \\ 
\hline
\begin{tabular}[c]{@{}l@{}}Learning rate \\decay schedule\end{tabular}                           & LinearWarmupCosineAnnealingLR                                                                                \\ 
\hline
\begin{tabular}[c]{@{}l@{}}Stopping criteria, \\and optimal model selection criteria\end{tabular} & \begin{tabular}[c]{@{}l@{}}training 300 epoches, \\no other model selection criteria\end{tabular}            \\ 
\hline
Data Augmentation                                                                                & \begin{tabular}[c]{@{}l@{}}RandZoom\\RandGaussianNoise\\RandGaussianSmooth\\RandScaleIntensity\end{tabular}  \\
\hline
\end{tabular}
\end{table}

\begin{table}[htbp]
\centering
\caption{Training Environment}
\begin{tabular}{|l|l|}
\hline
Item             & Configure                                      \\ \hline
Operating System & windows-10-10.0.19043-sp0                \\ \hline
Memory           & 190GB                                          \\ \hline
CPU              & Intel Xeon Gold 6142 M 2.60 GHZ * 16 \\ \hline
GPU              & NVIDIA A100-PCIE-80GB *2                           \\ \hline
\end{tabular}
\end{table}

\section{Results}
During this challenge, we make a 5-fold cross-validation experiment on the two-stage network to judge whether the training process can converge when using 3D CNN, eg.3D U-Net network mentioned in section2.3. The result is shown at table 5.1.3D convolution is able to converge and our proposed method outperforms 3D-UNet.

Our submitted docker image inference process can be summarized into three steps: (1) Coarse model locating ROI, (2) Fine mode segmentation on the ROI, (3) 3D-fixpoint method refinement the result.
The one-stage restoration method to obtain ROI coordinates is as follows:
  \begin{equation}
                    OriginalCoordinate=\frac{ResizedCoordinate*ResizedSpacing}{OriginalSpacing}
                        \end{equation}
Because of the objective inaccurate positioning problem, before restore the coordinate we add a fixed value to it, the value was set  as 2. Sliding window inference is performed on the cropped area, and the overlapping area between the sliding windows is set to 0, because the sampling patch method is according to the voxel, setting it to 0 will nearly not produce stitching artifacts. The area delimited by each sliding window will be transposed into the coronal plane, the axial plane, and the horizontal plane for inference. We do not activate the feature map of each plane separately, by contrast, we average them into one feature map before activating. 

In the validation phase we found that fine-tuning one part of  training data to improve performance will not improve the overall performance, the improvement of part data in the validation set will always be accompanied by the decline of another part of the data, we believe that the reason is the inconsistency of manual annotations. In order to solve such contradiction, we add the results of different models's outputs rather than average them. In the submitted docker image, stage two has two weight files. The first weight file is coming from under the strategy shown in the Table 4.2, and the second one is on the basis of the first weight file, fine-tune the 10 poor cases in the training set.

In order to eliminate outlier false positives caused by the background, the complete segmentation results will be sent to a 3D-fixpoint process for iteration. The difference between the 3D-fixpoint process and the previous fine model segmentation is that before 3D-fixpoint performs the same three-direction sliding window inference, it first extracts the maximum connected component according to the input voxels, and then expand the extracted boarder area. 3D-fixpoint process will exect 2 times and the expanded voxel num is set 5.

Finally, our two-stage method won the ninth place in the PARSE 2022 challenge in the testing phase. Part of results on the validation set is shown as Fig 5.1.

\begin{figure}
\centering
\includegraphics[width=0.5\linewidth]{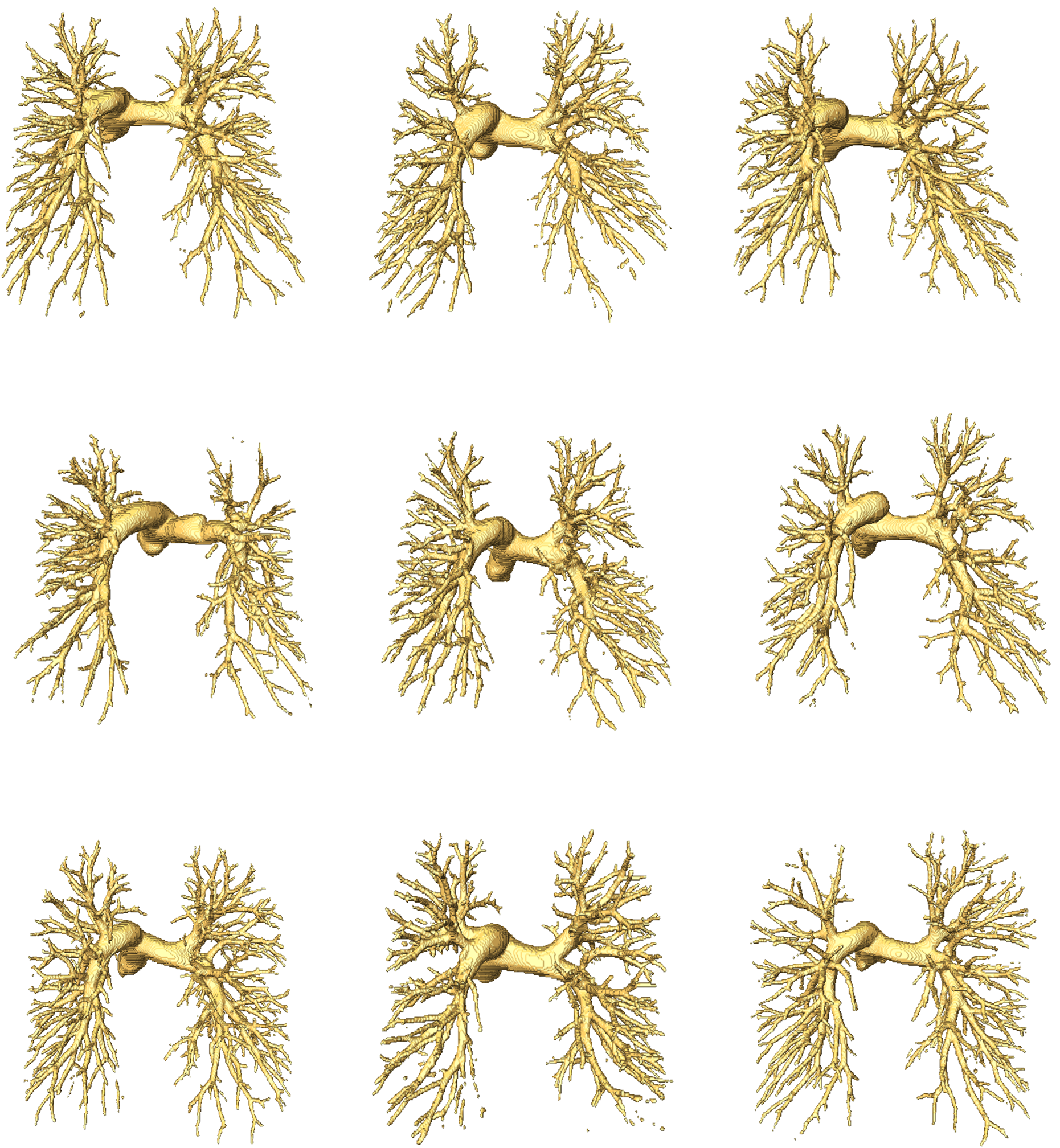}
\caption{9 cases of validation set}\label{fig5}
\end{figure}

\begin{table}
\centering
\caption{Five fold cross-validation on training set}
\begin{tabular}{|l|l|l|l|} 
\hline
        & Model                                                                & Dice                                                        & HD                                                            \\ 
\hline
Fold 1  & \begin{tabular}[c]{@{}l@{}}3D U-Net\\\textbf{ours}\end{tabular} & \begin{tabular}[c]{@{}l@{}}0.78\\\textbf{0.85}\end{tabular} & \begin{tabular}[c]{@{}l@{}}88.6\\\textbf{34.1}\end{tabular}   \\ 
\hline
Fold 2  & \begin{tabular}[c]{@{}l@{}}3D U-Net\\\textbf{ours}\end{tabular} & \begin{tabular}[c]{@{}l@{}}0.79\\\textbf{0.87}\end{tabular} & \begin{tabular}[c]{@{}l@{}}105.2\\\textbf{86.6}\end{tabular}  \\ 
\hline
Fold 3  & \begin{tabular}[c]{@{}l@{}}3D U-Net\\\textbf{ours}\end{tabular} & \begin{tabular}[c]{@{}l@{}}0.80\\\textbf{0.88}\end{tabular} & \begin{tabular}[c]{@{}l@{}}104.0\\\textbf{42.9}\end{tabular}  \\ 
\hline
Fold 4  & \begin{tabular}[c]{@{}l@{}}3D U-Net\\\textbf{ours}\end{tabular} & \begin{tabular}[c]{@{}l@{}}0.79\\\textbf{0.86}\end{tabular} & \begin{tabular}[c]{@{}l@{}}100.8\\\textbf{47.2}\end{tabular}  \\ 
\hline
Fold 5  & \begin{tabular}[c]{@{}l@{}}3D U-Net\\\textbf{ours}\end{tabular} & \begin{tabular}[c]{@{}l@{}}0.78\\\textbf{0.87}\end{tabular} & \begin{tabular}[c]{@{}l@{}}97.3\\\textbf{40.2}\end{tabular}   \\ 
\hline
Average & \begin{tabular}[c]{@{}l@{}}3D U-Net\\\textbf{ours}\end{tabular} & \begin{tabular}[c]{@{}l@{}}0.79\\\textbf{0.87}\end{tabular} & \begin{tabular}[c]{@{}l@{}}99.2\\\textbf{43.8}\end{tabular}   \\
\hline
\end{tabular}
\end{table}

\section{Discussion and Conclusion}

For the dataset of this challenge, we propose a multi-view, multi-stage and multi-window framework segmentation method to segment the pulmonary artery, one-stage for locating the pulmonary artery region, and two-stage for fine segmentation. In addition, we also adopt a fine-tune method according to the performance of the validation set to improve our submitted docker image performance. The model of each stage is based on 3D convolution, so we are at a disadvantage place in terms of memory and processing time. However, due to the inconsistency of manual labeling, such efforts are worthwhile, and the final result is clinically better than manual annotation. Our future research work includes hard sample mining, semi-supervised learning, registration, etc.
%
%
%
%

\bibliographystyle{splncs04}
\bibliography{samplepaper.bib}
\end{document}